\documentclass[conference]{IEEEtran}
\ifCLASSINFOpdf
\else
\fi

\usepackage{graphicx}
\usepackage{placeins}

\usepackage{textcomp} % for degree symbol

\begin{document}
\title{Multi-Contrast CT Imaging with a Prototype Spatial-Spectral Filter}

\author{\IEEEauthorblockN{Matthew Tivnan, Wenying Wang, J. Webster Stayman}
\IEEEauthorblockA{Department of Biomedical Engineering\\
Johns Hopkins University\\
720 Rutland Ave., Baltimore MD, 21205 \\
Email: web.stayman@jhmi.edu}}


%Telephone: (800) 555--1212, Fax: (888) 555--1212}


\maketitle

\begin{abstract}
Spectral CT has great potential for a variety of clinical applications due to the improved material discrimination with respect to conventional CT. Many clinical and preclinical spectral CT systems have two spectral channels for  dual-energy CT using strategies such as split-filtration, dual-layer detectors, or kVp-switching. However, there are emerging clinical imaging applications which would require three or more spectral sensitivity channels, for example, multiple exogenous contrast agents in a single scan. Spatial-spectral filters are a new spectral CT technology which use x-ray beam modulation to offer greater spectral diversity. The device consists of an array of k-edge filters which divide the x-ray beam into spectrally varied beamlets. This design allows for an arbitrary number of spectral channels; however, traditional two-step reconstruction-decomposition schemes are typically not effective because the measured data for any individual spectral channel is sparse in the projection domain. Instead, we use a one-step model-based material decomposition algorithm to iteratively estimate material density images directly from spectral CT data. In this work, we present a prototype spatial-spectral filter integrated with an x-ray CT test-bench. The filter is composed of an array of tin, erbium, tantalum, and lead filter tiles which spatially modulate the system spectral sensitivity pattern. After the system was characterized and modeled, we conducted a spectral CT scan of a multi-contrast-enhanced phantom containing water, iodine, and gadolinium solutions. We present the resulting spectral CT data as well as the material density images estimated by model-based material decomposition. The calibrated system model is in close agreement with the measured data, and the reconstructed material density images match the ground truth concentrations for the multi-contrast phantom. These preliminary results demonstrate the potential of spatial-spectral filters to enable multi-contrast imaging and other new clinical applications of spectral CT.
\end{abstract}


% \fi
\IEEEpeerreviewmaketitle

\section{Introduction}

Spectral CT is a term used to describe any CT system which incorporates projection data acquisition with varied x-ray sensitivity spectra to allow for both spatial reconstruction as well as basis material density estimation. Dual-energy CT is an existing subcategory of spectral CT which is already being successfully utilized in clinical settings. A recent comprehensive review of the modern state of of the technology and its applications \cite{mccollough2015dual} enumerates important new clinical tasks made possible by dual-energy CT including automated bone removal in CT angiography \cite{buerke2009dual}, virtual monochromatic imaging \cite{yu2011virtual}, and virtual non-contrast imaging \cite{ferda2009assessment}. 

Greater diversity of spectral sensitivity has the potential to enable entirely new clinical imaging applications. For example, a multi-contrast protocol for simultaneous multi-phasic liver imaging with spectral CT is already under development \cite{muenzel2017simultaneous}. With the goal of imaging multiple biologically safe contrast agents (such as iodine and gadolinium), future systems will need to move beyond dual-energy CT to designs which incorporate three of more sensitivity spectra into a single data acquisition.

\begin{figure}[ht]
    \centering
    % trim {left, bottom, right, top}
    \includegraphics[trim={0cm, .4cm, 0cm, 1cm},clip,width=0.42\textwidth]{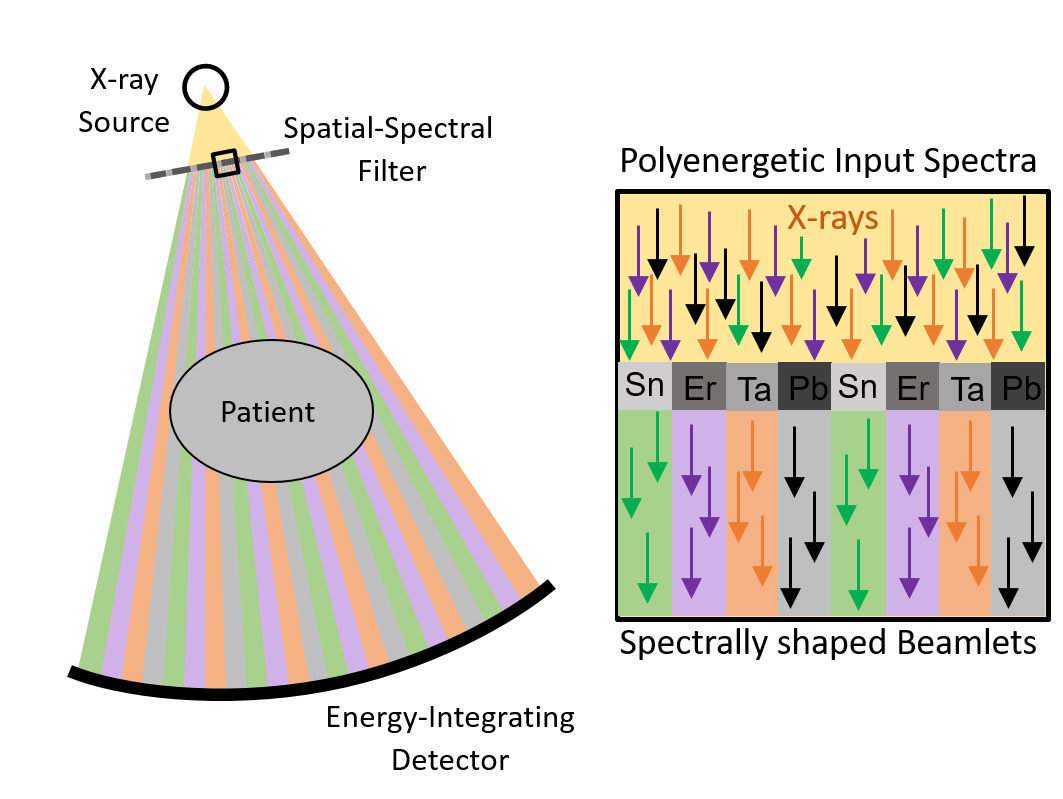}
    \caption{Diagram of
    spatial-spectral filters for spectral CT. A tiled array of k-edge filters (e.g. tin, erbium, tantalum, lead) which modulate the spectra of the incident x-rays. The resulting projection data have varied spectral sensitivity.}
    \label{fig:cartoon}
\end{figure}

One possible solution would be the use of direct photon-counting detectors which are capable of photon-energy discrimination. This technology shows great promise for spectral CT in the future but currently faces its own set of limitations and challenges \cite{taguchi2013vision}. We focus on a new technology which is compatible with current energy-integrating detectors called spatial-spectral filters which are shown in Figure \ref{fig:cartoon}. In previous work, we have modeled and optimized the filter design \cite{tivnan2019optimized}%\cite{tivnan2019best} \cite{tivnan2019physical}
. Using those simulation results as guidance for the device design, we have constructed a prototype spatial-spectral filter in an integrated x-ray CT test bench system. 

In this work, we present the prototype spatial spectral filter as well as specialized system calibration and data processing procedures required to use the device for spectral CT. We investigate the efficacy of the spatial-spectral CT system using a multi-contrast phantom containing iodine and gadolinium solutions. We present the spectral CT projection data as well as the basis material density images reconstructed using a direct one-step model-based material decomposition algorithm.

\section{Methods}

\subsection{Spatial-Spectral Filter Prototype}

\begin{figure}[ht]
    \centering
    % trim {left, bottom, right, top}
    \includegraphics[trim={1.5cm, 0cm, 0cm, 0.3cm},clip,width=0.23\textwidth]{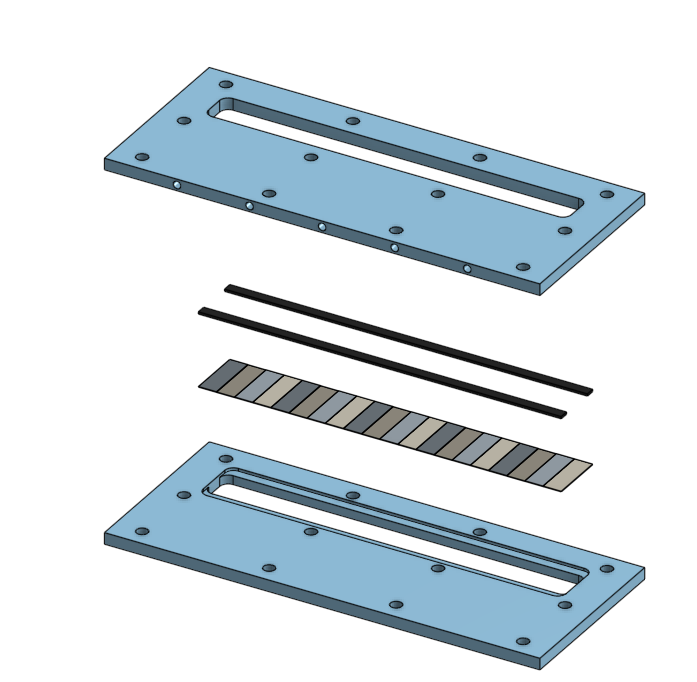}
    \includegraphics[trim={1cm, 0cm, 1.5cm, 1cm},clip,width=0.15\textwidth]{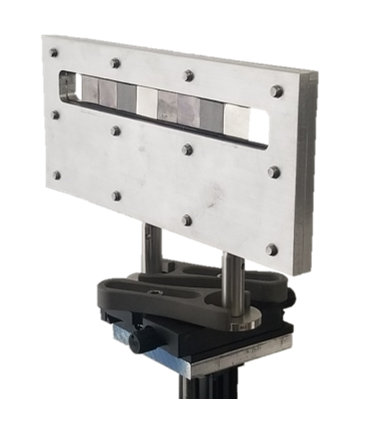}
    \caption{Mechanical design of spatial-spectral filter prototype. The tin, erbium, tantalum, and lead filter tiles clamped in an aluminum frame. }
    \label{fig:filter_cad}
\end{figure}

The prototype spatial-spectral filter consists of tin, erbium, tantalum, and lead filter tiles (ordered as such) which are 20~mm wide, 30~mm high and 0.250~mm thick with the exception of the tantalum tiles which are 0.127~mm thick. The relative position of the filter tiles is clamped in place by an aluminum frame as shown in Figure \ref{fig:filter_cad}. The entire assembly is integrated into an x-ray CT test bench. It is mounted in front of the x-ray source onto a linear actuator which can translate the device laterally. This axis of motion allows for repositioning throughout the CT scan and corresponding view-dependent modulation of the system spectral sensitivity. 
 
\FloatBarrier

\vspace{-3mm}

\subsection{Data Processing and Calibration}

The spatial-spectral filter roughly divides the projection data into four spectral channels associated with the filter materials. Each of these channels is individually sparse in the projection domain, taking up approximately 25\% of the CT dataset. With this incomplete CT data, it would be difficult to reconstruct images for each spectral channel individually which would be a pre-requisite for material decomposition in the image domain. Projection-domain decomposition is also challenging without any overlap between the spectral channels' acquisition geometries. Instead, we estimate material densities directly from the spectral CT data with a one-step model-based material decomposition algorithm \cite{tilley2018model} \cite{tivnan2020preconditioned}. This algorithm involves a poly-energetic forward model for x-ray attenuation that properly considers the mass attenuation spectra for each of the basis materials, as well as the models for the system geometry and sensitivity spectra. The data log-likelihood term assumes a multivariate Gaussian noise model, and the regularization term is a quadratic smoothness penality with cross-material penalties as described in \cite{wang2019generalized}.

System spectral sensitivity is affected by the source fluence spectra, the source-side filtration, and the detector energy-dependent sensitivity. For model-based material decomposition, it is critical to have an accurate model for system spectral sensitivity. The spectral model was calibrated by fitting aluminum and tungsten source filtration to the measured transmissivity spatial-spectral filter.

Small gaps between the tiles result in an unfiltered beam and a corresponding increase in x-ray fluence. We have shown in previous work that reconstruction with filter tile gaps is possible by incorporating the gaps into the system spectral sensitivity model. However, in this work, we simply set the statistical weights to zero around the interface between filter tiles, thereby ignoring this data and avoiding the more complex calibration required to model the gaps.

\subsection{Spatial-Spectral CT  with a Multi-Contrast Phantom}

\begin{figure}[ht]
    \centering
    % trim {left, bottom, right, top}
    \includegraphics[trim={0cm, 0cm, 0cm, 4cm},clip,width=0.48\textwidth]{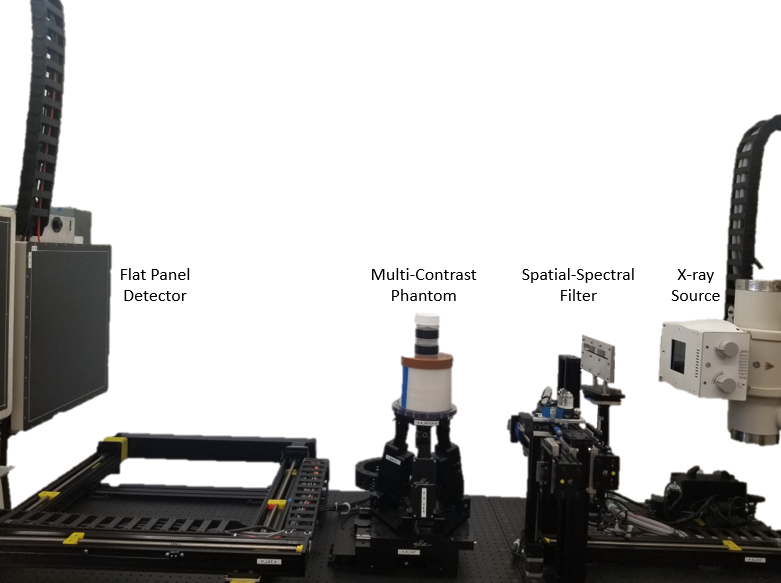}
    \caption{X-ray test bench outfitted for spectral CT with a spatial-spectral filter}
    \label{fig:bench}
\end{figure}

With the clinical goal of multi-contrast imaging, we conducted a spectral CT experiment with the spatial-spectral filter x-ray test bench system in Figure~\ref{fig:bench}. The subject of the scan is a multi-contrast phantom consisting of a 65~mm diameter phantom cylindrical water tank and 10~mm diameter vials containing various concentrations of iodine and gadolinium solutions distributed as shown in Figure~\ref{fig:images}. The source-to-axis distance was 910~mm and the source-to-detector distance was 1920~mm. The spatial-spectral filter was positioned 380~mm from the source and it was translated at a speed of 2.66~mm per view. The flat-panel detector has a caesium iodide scintillator thickness of 0.60~mm  and a pixel spacing of 0.278~mm in both the row and column directions. 

\FloatBarrier

The x-ray technique was 120~kVp with one 40~ms pulse per frame at a current of 30~mA. Air and offset scans were collected with no filtration and no phantom in place. Then, a filter calibration scan was collected with the view-dependent filtration from the spatial-spectral filter, but no phantom in place. After standard air and offset corrections, the spectral calibration parameters were tuned such that the measured transmissivitiy of all four filter tile materials matched the model used for reconstruction for the known filter tile thicknesses. Projection data were acquired over 360° using 360 equally spaced view angles.

We ran 1000 iterations of the material decomposition algorithm. The voxelized representation of the object consisted of 88 voxels with 1~mm spacing. We used quadratic regularization to encourage smooth image results and we used a cross-material penalty to reduce cross-talk artifacts between basis material channels.

\section{Results}

\subsection{Calibrated Spectral Sensitivity}

\begin{figure}[ht]
    \centering
    % trim {left, bottom, right, top}
    \includegraphics[trim={8mm, 0cm, 1cm, 0cm},clip,width=0.49\textwidth]{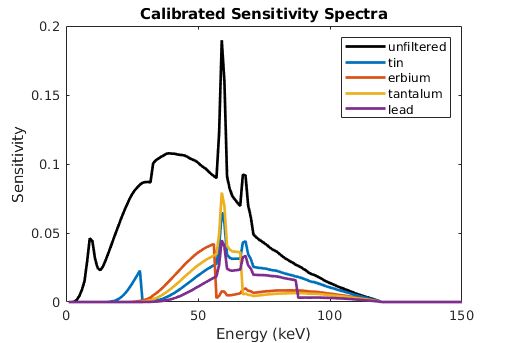}
    \caption{Calibrated spectral sensitivity for each filter material.}
    \label{fig:spectra}
\end{figure}

The transmissivity of the spatial-spectral filter was 38.59\% for tin, 20.82\% for erbium, 30.85\% for tantalum, and 23.97\% for lead. Based on these levels, the spectral calibration output parameters are -0.0378~mm of aluminum filtration and 0.1436~mm of tungsten filtration relative to the unfiltered SPEKTR model. The output spectra are shown in Figure~\ref{fig:spectra}. 

In these spectra, we can see the k-edges at 29.20~keV for tin, 57.48~keV for erbium, 67.43~keV for tantalum, and 88.01~keV for lead. This type of k-edge filtration leads to diverse sensitivity spectra which help to separate the signals from water, iodine, and gadolinium, based on their respective mass attenuation spectra.

\subsection{Multi-Contrast Imaging Experiment}

\begin{figure}[ht]
    \centering
    % trim {left, bottom, right, top}
    \includegraphics[trim={0cm, .5cm, 0cm, .5cm},clip,width=0.48\textwidth]{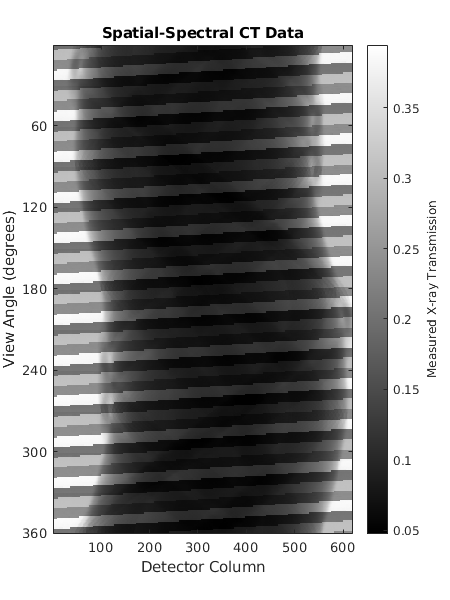}
    \caption{Spatial-spectral CT data of a multi-contrast phantom scan.}
    \label{fig:data}
\end{figure}

The spectral CT scan of the multi-contrast phantom acquired with the spatial-spectral filter is shown in Figure \ref{fig:data}.  The linear motion of the filter throughout the scan results in the diagonal stripes. Note that any individual spectral channel corresponding to a certain filter material does not make up a complete CT dataset. We only achieve sufficient spatial and spectral sampling by considering all the data together. In total, 3\% of the projection data were identified as part of the interface region between filter tiles and were disregarded in the statistical reconstruction. In Figure \ref{fig:data}, those projections have been replaced via interpolation between neighboring views.

We can see the projection of the water tank as well as the lower-contrast projection signal of the iodine and gadolinium inserts. Based on the sensitivity spectra for a given projection, the same mass-length of a certain material will result in a different contrast. Only by considering this data together with a model for the system geometry and projection-dependent spectral sensitivity can we estimate the basis material densities.

\subsection{Model-Based Material Decomposition}

\begin{figure*}[ht]
    \centering
    % trim {left, bottom, right, top}
    \includegraphics[trim={-1cm, .5cm, -3cm, 0cm},clip,width=0.78\textwidth]{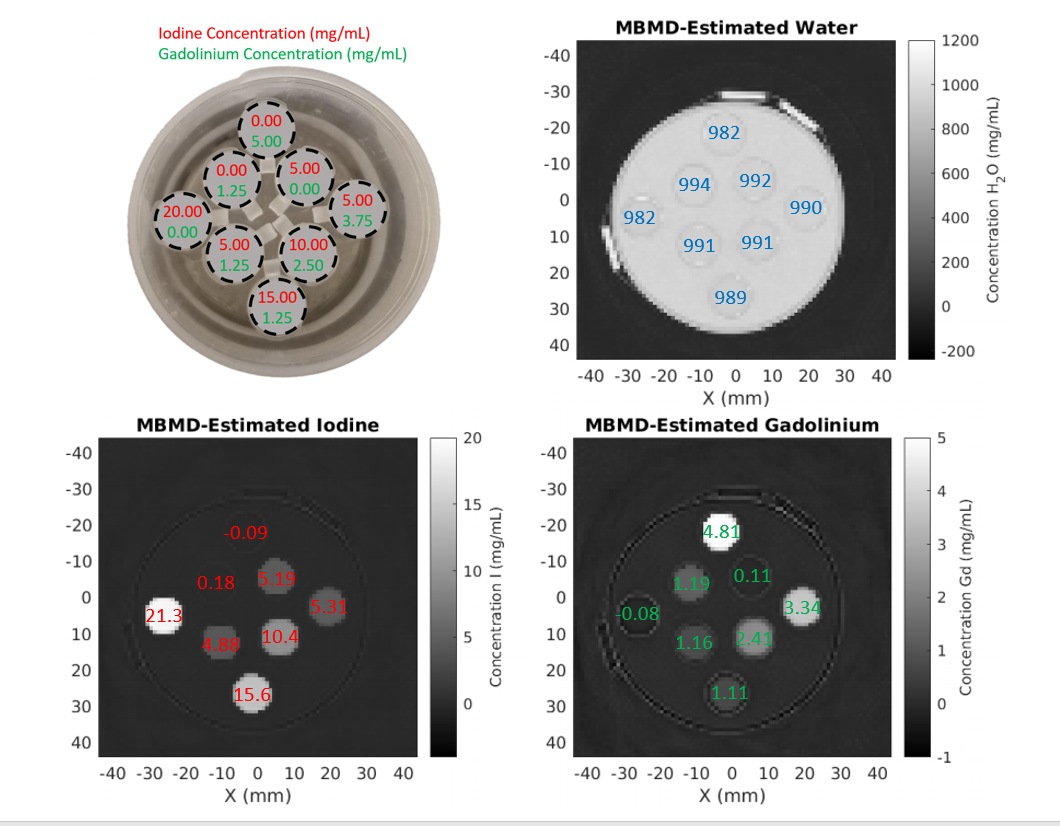}
    \caption{Multi-contrast phantom with vials containing iodine and gadolinium solutions at various mixtures in water. Material density images estimated via model-based material decomposition.}
    \label{fig:images}
\end{figure*}

The output basis material density images from the model-based material decomposition are shown in Figure~\ref{fig:images}. The figure also shows overlays of ROI mean values which demonstrate a very close match to the ground truth concentrations for each material. 

The water channel shows a relatively flat density with the exception of the plastic vials and the plastic rectangular fiducials on the outside of the tank, which are used to verify the orientation. The mean value of the flat response in the water tank is around 992~mg/mL which is close to the ground truth value of 1000~mg/mL. Some of the attenuation is falsely attributed to an elevated baseline in the gadolinium channel of around 0.78~mg/mL for regions of the phantom that should only contain water. This likely indicates some errors in the spectral calibration.

For vials containing a mixture of both gadolinium and iodine, the resulting decomposition appears to over-estimate the iodine and under-estimate the gadolinium. This is demonstrative of the strong anti-correlation between materials which poses a great challenge for material decomposition problems.Again, these errors could be a bias due to small errors in the modeled system sensitivity spectra. Finally, are some artifacts the material boundaries which could be caused by cross-talk from the quadratic regularization penalty or by bias from unmodeled spectral effects.

%These artifacts are known to be a result of the regularization. When we enforce a smoothing penalty on an edge in the water channel, the data likelihood term encourages increased density in other channels. In this case, the gadolinium channel appears to be the most strongly effected. We attempted to reduce these errors with a shift-invariant cross-material penalty, however, the effect is also known to be shift variant, so a shift-variant penalty may lead to improvements in the future. 

\newpage
\section{Conclusion}

\vspace{-1mm}

The hardware designs, data processing procedures, and experimental results presented here show that spatial-spectral filters show promise as a strategy for spectral CT with energy integrating detectors. While spatial-spectral filters yield sparse projection data for each spectral channel, these results shown that we can use one-step material decomposition algorithms to reconstruct basis material density images using the proper calibration and physical models. Morever, using these spatial-spectral filters we were able to conduct a multi-contrast imaging experiment using spectral-modulation of the x-ray beam without the need for energy-discriminating detectors.

Spectral CT shows promise for many clinical applications. As we develop new technologies which increase the number of spectral channels, we have the potential enable entirely new clinical studies. For example, three or more spectral channels enables three or more basis materials which is the fundamental requirement for multi-contrast imaging and quantification (one of the basis materials being water). Creative new clinical studies for multiple contrast agents are already under development for liver imaging, and it is reasonable to expect more new applications in the future. 

In ongoing work, we seek to refine procedures to calibrate gaps and edges between filters for full data utilization. We will also aim to investigate combining spatial-spectral filters with other techniques like kVp-switching or energy-discriminating detectors to explore high-sensitivity spectral CT.

\small

\section*{Acknowledgements}

This work was supported, in part, by NIH grants R21EB026849.

\newpage
% \section*{References}
\bibliography{report}
\bibliographystyle{plain} % makes bibtex use spiebib.bst

\end{document}